# On the internal gravity waves in the stratified ocean with shear flows

**Vitaly V. Bulatov\*, Yury V. Vladimirov**

**Institute for Problems in Mechanics**

**Russian Academy of Science**

Address: Pr. Vernadskogo 101 - 1, Moscow, 119526, Russia

\*Correspondence: bulatov@index-xx.ru

fax: +7-499-739-9531

**Abstract.**

In this paper, we consider a fundamental problem of describing the dynamics of internal gravity waves in the stratified ocean with shear flows. We develop an asymptotic representation of the wave fields in terms of the Green's functions. We explore the far field of the internal gravity waves generated by disturbing sources, and propose asymptotic algorithms for calculating its dynamics.

**Problem formulation.**

There are always flows with the vertical shear in a real ocean. These flows have a significant impact on the internal gravity waves dynamics. For real oceanic conditions on frequent occasions we have to analyze the internal gravity waves that propagate against the background of mean frictions with a vertical velocity shift, at that the vertical velocity variation comes up to tens of sm/s and m/s, that is, of the same order as the maximum velocities of internal gravity waves. It is evident that such flows shall have a substantial impact on the propagation of internal gravity waves. If the scope of horizontal changes of flows is much higher than the lengths of internal waves, and the scope of temporal variability is much larger than the periods of internal waves, then a natural mathematical model will be the case of stationary and horizontal uniform average shear flows. To elucidate the mechanism of interaction between the shear lows and internal gravity waves, we consider only the basic models. We assume that the shear flow changes only vertically. Generalization of the results can give a qualitative understanding of the complex processes occurring in the real ocean. The geophysical importance of this problem due to the fact that the largest vertical and horizontal shifts in the real ocean associated with the interaction of shear flows and internal gravity waves. It is also necessary to note the complexity of the mathematical solutions to these problems [1-9].

Let $\mathbf{V}(z) = (V_1(z), V_2(z))$ be the shear flow on the horizon $z$. Linearizing the hydrodynamics equation with regard for non-excitation state, where $U_1 = V_1(z)$, $U_2 = V_2(z)$, $W = 0$, $r = r_0(z)$, $p = p_0(z) = -\int r_0 dz$, we can obtain a system of equations [10-15]

$$r_0 \frac{DU_1}{Dt} + \frac{\partial p}{\partial x} = 0, \quad r_0 \frac{DU_2}{Dt} + \frac{\partial p}{\partial y} = 0, \quad r_0 \frac{DW}{Dt} + \frac{\partial p}{\partial z} + rg = 0$$

$$\frac{\partial U_1}{\partial x} + \frac{\partial U_2}{\partial y} + \frac{\partial W}{\partial z} = 0 \qquad (1)$$

$$\frac{\partial r}{\partial t} + W \frac{\partial r_0}{\partial z} = 0$$

$$\frac{D}{Dt} = \frac{\partial}{\partial t} + V_1(z)\frac{\partial}{\partial x} + V_2(z)\frac{\partial}{\partial y}$$

Making use of the Boussinesq approximation, that is, assuming here $\frac{\partial r_0}{\partial z} = -\frac{N^2(z)}{g}$ ($N^2(z)$

- Brent-Vaisala frequency, main parameter of internal gravity wave dynamics in real ocean [12-14]) and $r_0 = 1$, and excluding after that the functions $U_1, U_2, p, r$, we get the equation for the vertical velocity

$$L^*W = 0 \qquad (2)$$

$$L^* = \frac{D^2}{Dt^2}\Delta_3 - \frac{D}{Dt}\left(\frac{\partial^2 V_1(z)}{\partial z^2}\frac{\partial}{\partial x} + \frac{\partial^2 V_2(z)}{\partial z^2}\frac{\partial}{\partial y}\right) + N^2(z)\Delta,$$

$$\Delta = \frac{\partial^2}{\partial x^2} + \frac{\partial^2}{\partial y^2}, \qquad \Delta_3 = \Delta + \frac{\partial^2}{\partial z^2}.$$

The boundary conditions for (2) shall be taken in the "rigid-lid" approximation

$$W = 0 \quad (z = 0, -H). \qquad (3)$$

**Problem solutions and main results.**

First of all consider the plane waves, that is, the solutions in the form of

$$W = \exp(i(lx + my - wt))j(z) \qquad (4)$$

Substituting this formula into (1), we obtain the equation for $j$ (the Taylor-Goldstein equation) and the boundary problem

$$L_0 j = (w-kF)^2 \frac{\partial^2 j}{\partial z^2} + \left\{ k^2 [N^2(z) - (w-kF)^2] + k \frac{\partial^2 F}{\partial z^2}(w-kF) \right\} j = 0, \qquad (5)$$

$$j = 0 \quad (z = 0, -H).$$

Here $l = k\cos y$, $m = k\sin y$, $F = F(z) = V_1(z)\cos y + V_2(z)\sin y$ is the flow velocity component $\mathbf{V}(z)$ in the direction of the wave propagation (4).

Consider some peculiarities of the spectral problem (5), where the spectral parameter is $w$. Next, it will be convenient after substituting the unknown function

$$j(z) = (w - kF(z))u(z) \qquad (6)$$

to go to the problem

$$L_1 u = \frac{d}{dz}\left[ (w-kF)^2 \frac{\partial u}{\partial z} \right] + k^2 [N^2(z) - (w-kF)^2] u = 0. \qquad (7)$$

Given there are average shearing flows the internal waves while interacting with those flows can exchange energy with them, that is, oscillations (4) can be exponentially decaying (at $\mathrm{Im}\, w < 0$) or rising (at $\mathrm{Im}\, w > 0$). To avoid this it is required that the vertical gradient of average flows be not too large as compared to the buoyancy frequency. It is sufficient to ask that the Mails stability condition is fulfilled, that is, at all $z$ the inequation holds [1,12-14]

$$\left( \frac{\partial V_1(z)}{\partial z} \right)^2 + \left( \frac{\partial V_2(z)}{\partial z} \right)^2 \leq 4N^2(z) \qquad (8)$$

Let's prove that if the inequation (8) holds, then the spectral problem (7) may not have complex eigenvalues $w = w_r + iw_i$. To this end note that for any functions $f(z)$ and $g(z)$ going to zero at $z = 0, -H$ there is a valid inequality

$$\int_{-H}^{0} gL_1 f \, dz = \int_{-H}^{0} k^2 (N^2(z) - \Phi^2) fg \, dz - \int_{-H}^{0} \Phi^2 \frac{\partial f}{\partial z} \frac{\partial g}{\partial t} dz \qquad (9)$$

$$\Phi = w - kF(z).$$

Now, let $f(z)$ be a solution of the boundary problem (7) at $w = w_r + iw_i$, assume $Q = \sqrt{\Phi}\, f(z)$ and $g = \Phi^{-1/2} Q^*(z)$, where $Q^*(z)$ is the complex conjugate function to $Q(z)$. Then the formula (5.2.9) becomes zero, and its right-hand member can be written in the form of

$$\int_{-H}^{0} k^2 \left[ \frac{N^2(z)}{\Phi} - \Phi \right] |Q|^2 dz - \frac{1}{4} \int_{-H}^{0} \frac{1}{\Phi} \left( \frac{\partial \Phi}{\partial z} \right)^2 |Q|^2 dz +$$

$$+ \frac{1}{2} \int_{-H}^{0} \left( \frac{\partial \Phi}{\partial z} \right) \frac{d}{dz} (QQ^*) dz - \int_{-H}^{0} \Phi \left| \frac{dQ}{dz} \right|^2 dz.$$

The imaginary part of the formula is equal to

$$- w_i \left[ k^2 \int_{-H}^{0} \left[ N^2(z) - \frac{1}{4} \left( \frac{\partial F}{\partial z} \right)^2 \right] |\Phi|^{-2} dz + \int_{-H}^{0} \left( k^2 |Q|^2 + \left| \frac{\partial Q}{\partial z} \right|^2 \right) dz \right].$$

Because

$$\left( \frac{\partial F}{\partial z} \right)^2 = \left( \frac{\partial V_1(z)}{\partial z} \cos y + \frac{\partial V_2(z)}{\partial z} \sin y \right)^2 \leq \left( \frac{\partial V_1(z)}{\partial z} \right)^2 + \left( \frac{\partial V_2(z)}{\partial z} \right)^2 \leq 4 N^2(z),$$

the formula in square brackets is positive and therefore it is essential that the condition $w_i = 0$ is fulfilled.

Thus, if the Miles stability condition (8) is met then the spectral problem (5) (or (7)) has no complex eigenvalues $w$. It can be proved that there are two sets of real eigenvalues $w$. In the first one $w_n$ are increasing and work for $kF_+ = \min_z kF(z)$, in the second set the eigenvalues $w_n$ are decreasing and work for $kF_+ = \max_z kF(z)$. We shall enumerate the first set with negative values and the second one with positive values; $|n|$ is the number of sign variations of the eigenfunctions $j_n$.

Note the qualitative difference between the behavior of eigenfunctions $j_n(z)$ at $|n| \to \infty$ in the event that there are flows or there are none. If there are no flows then the equation for $j_n$ is given by

$$w^2 \frac{\partial^2 j(z,k)}{\partial z^2} + k^2(N^2(z) - w^2) j(z,k) = 0,$$

$w_n \to 0$ at $n \to \infty$, and the eigenfunctions $j_n(z)$ become ever more oscillating and do not tend to any limit. On the other hand, if there are flows present and, for instance, $n \to \infty$, $w_n \to kF_+ = \max_z kF(z)$, the equation (7) is working for the limit equation

$$\frac{d}{dz}\left[(F_+ - F(z))^2 \frac{\partial u}{\partial z}\right] + \left[N^2(z) - k^2(F_+ - F(z))^2\right] u = 0 \tag{10}$$

and the eigenfunctions $u_n(z)$ at any fixed $z$, for which $F(z) \neq F_+$, and $n \to \infty$ are working for the solution $u(z)$ of this equation.

The Green's function in presence of average shearing flows satisfies the equation

$$L^* G(t, r, z, z_0) = d(t) d(z - z_0) d(x) d(y) \tag{11}$$

where $L^*$ is the operator (2), and it holds for zero boundary conditions (3) and initial conditions

$$G \equiv 0 \quad (t < 0) \tag{12}$$

As in the case of no flows, when using the Fourier method we obtain

$$G = \frac{1}{(2p)^3} \int\int_{-\infty}^{\infty} e^{i(lx+my)} dl\, dm \int_{-\infty+ie}^{\infty+ie} e^{-iwt} j(w, l, m, z, z_0) dw, \tag{13}$$

where $j$ is the solution of the equation and boundary problem

$$L_0 j = -d(z - z_0); \quad j = 0 \quad (z = 0, -H) \tag{14}$$

and $L_0$ is the Taylor-Golstein operator (5). At $\text{Im} w \neq 0$ the solution of this problem is unique because $L_0$ has no complex eigenvalues.

In the case of no flows we solve the equation similar to (14) by expanding $d$-function in series on eigenfunctions of the operator $L = w^2 \frac{\partial^2}{\partial z^2} + k^2(N^2(z) - w^2)$. However, where there are flows the eigenfunctions are not only non-orthogonal, but also non-complete, and to construct the solution (14) it is necessary to use other mode.

Let us set $v_1(z, w)$ and $v_2(z, w)$ as the solutions of the equation $L_0 v = 0$, which are going to zero respectfully at $z = 0$ and $z = -H$. Then the solution $j(w, l, m, z, z_0)$ of the boundary problem in (14) is given by

$$j(w, l, m, z, z_0) = \begin{cases} -\dfrac{v_1(z, w) v_2(z_0, w)}{(w - F(z_0))^2 W} & z > z_0 \\ -\dfrac{v_1(z_0, w) v_2(z, w)}{(w - F(z_0))^2 W} & z < z_0 \end{cases}, \tag{15}$$

where $W = W(w) = \frac{\partial v_1}{\partial z} v_2 - v_1 \frac{\partial v_2}{\partial z}$ is the Wronskian of the functions $v_1(z, w)$ and $v_2(z, w)$.

Let's analyze the behavior of $v_1$, $v_2$, Wronskian $W$ and $j(w, l, m, z, z_0)$ as the functions $w$. If $Jm w \neq 0$, then at $\frac{\partial^2 j}{\partial z^2}$ the coefficient in (5) is not going to zero at whatever $z$ and hence, the solutions $v_1$ and $v_2$ for the equation are regular at all $z$ and are the analytical functions $w$. Values $w$ at which $W$ goes to zero are eigenvalues of the operator $L_0$. Since the Miles stability condition is assumed to have been fulfilled this operator has no complex eigenvalues, $W$ has no complex zeroes and $j$ is analytic at any complex $w$.

If $w$ is real, but $w < kF_- = k \min_z F(z)$ or $w > kF_+ = k \max_z F(z)$, then the coefficient at $\dfrac{\partial^2 j}{\partial z^2}$ in (5) still is not going to zero at whatever $z$ and the functions $v_1$, $v_2$ are the analytical functions $w$. However, the Wronskian $W$ already can go to zero; its zeroes are eigenvalues $w_n$. Deductions $j$ at $w = w_n$ are expressed by $\left.\dfrac{\partial W}{\partial w}\right|_{w=w_n}$, that is, by

$$d_n = \frac{\partial}{\partial w}\left[\frac{\partial v_2(-H)}{\partial z} v_1(-H)\right]\bigg|_{w=w_n} = \left.\frac{\partial j_n}{\partial w}\frac{\partial j_n}{\partial z}\right|_{z=-H}.$$

For $d_n$ the below formula is true

$$d_n = 2\int_{-H}^{0} (w_n - kF)\left\{\left[\frac{d}{dz}\frac{j_n}{(w_n - kF)}\right]^2 + \frac{k^2 j_n^2}{(w_n - kF)^2}\right\} dz. \qquad (16)$$

Finally, at $kF_- < w < kF_+$ we have values $z$, at which the coefficient $(w - kF)^2$ at $\dfrac{\partial^2 j}{\partial z^2}$ in (5) goes to zero. This formula $z$ is the ramification point for the solution of the equation (5). Thus, the interval $kF_- < w < kF_+$ is the cut for functions $v_1$, $v_2$ and the Wronskian $W$. If $w$ is inside this interval, then the limits $j(w+ie)$ and $j(w-ie)$ (where $e \to 0$ and $j$ are given by (15)) are complex joined and differ from each other. At that we can prove that $W$ is not going to zero on the upper or lower banks of this cut, that is, all zeroes of $W$ are completed by the above indicated series $w_{-n}$ and $w_n$.

Let's calculate the integral over $w$ in (13). At $t < 0$ when diverting the contour of integration on infinity in the upper semi-plane we get a zero. At $t > 0$ the integration contour is closed in the lower semi-plane and the integral is reduced to a sum of deductions and integral over the cut

$$\Gamma = \int_{-\infty+ie}^{\infty+ie} j e^{-iwt} dw = 2pi \sum_{n=-\infty}^{\infty} e^{-iw_n t} \frac{j_n(z)j_n(z_0)}{d_n(w_n - kF(z_0))^2} + \Gamma_m \tag{17}$$

Here the summation goes over eigenfunctions of the operator (5), that is, over a discrete spectrum, and $\Gamma_m$ is the integral over the cut, that is, over the continuous spectrum of this operator:

$$\Gamma_m = 2i \int_{kF_-}^{kF_+} e^{-iwt} \operatorname{Im} j(w+i0, l, m, z, z_0) dw.$$

Where there are no flows the formula similar to (17) is given by

$$g = \frac{1}{4p^2} \sum_{n=-\infty}^{\infty} \frac{1}{2ik^2} e^{-iw_n t} w_n(k) j_n(z,k) j_n(z_0,k) \tag{18}$$

where at $n < 0$ we have set $w_n(k) = -w_{-n}(k)$, $j_n(z,k) = j_{-n}(z,k)$.

The qualitative difference between the formulas (17) and (18), apart from the integral existing in (17) over the continuous spectrum $\Gamma_m$, lies in the nature of the discrete spectrum convergence. In the formula (18) the eigenvalues $w_n$ behave like $n^{-1}$, and the series (18) converges conditionally, at the cost of oscillations $j_n(z,k) j_n(z_0,k)$. In the event there are flows and, for example, $n \to \infty$, $w_n$ is tending to $kF_+$, the functions $j_n(z)$ and $j_n(z_0)$ go to the solutions of the equation (10), and the convergence of the series (17) is provided by fast increasing $d_n$. It is possible to demonstrate that $d_n \approx n^3$ at $n \to \infty$, that is, the series (17) converges absolutely.

Note, by the way, that the expansion of functions $\Gamma$ (17) includes an integral over the continuous spectrum, and it follows that for this expansion missing is the series on eigenfunctions, that is, the system of eigenfunctions $j_n(z)$ is not complete.

The eigenfunctions $j_n(z)$ in (17) depend upon $l, m$. It's easy to see that

$$j_{-n}(z,-l,-m) = j_n(z,l,m), \quad w_{-n}(-l,-m) = -w_n(l,m).$$

Taking into account these relationships and integrating (5.2.17) over $l$ and $m$, we obtain the following formula for the Green's function:

$$G = \sum_{n=1}^{\infty} G_n(t,x,y,z,) + G_m(t,x,y,z,),$$

$$G_n = \frac{1}{2p^2} \text{Im} \int_{-\infty}^{\infty}\int_{-\infty}^{\infty} \exp(i(lx+my-w_n t))\frac{j_n(z)j_n(z_0)dl\,dm}{d_n(w_n - kF(z_0))^2} \tag{19}$$

is the $n$-th mode, and

$$G_m = \frac{1}{(2p)^2} \iint \exp(i(lx+my))\Gamma_m(t,l,m,z,z_0)dl\,dm$$

is the integral over the continuous spectrum.

**Asymptotic representations of far wave fields.**

Let's analyze the Green's function $G$ asymptotics in the far region, that is, at $t, x, y \to \infty$; $\frac{x}{t}, \frac{y}{t} = 0(1)$. It may be shown that under these conditions the integral $G_m$ is decaying at $t \to \infty$ faster than any exponent $t$. With this in mind in the following we shall omit this term and analyze the asymptotics in the far region of every $G_n$ mode.

Assume $x = at; y = bt$. Then the phase function in (19) shall be written in the form

$$\Phi = t[l a + mb - w_n(l,m)] \tag{20}$$

and the asymptotics $G_n$ at larger $t$ shall be defined, firstly, by exceptional points for which $\Phi$ and the non-exponential factor in (19) lose their analytic property as functions $l, m$, and, secondly, by stationary points $\Phi$.

Such an exceptional point is the value $l = m = 0$, which is a conal exceptional point for $w_n(l, m)$. Indeed, let's assume $l = k\cos y$, $m = k\sin y$. As we can see below, even at smaller $k$ the function $w_n(l, m)$ is given as $w_n = kx_{0n}(y) + O(k^3)$. The non-exponential factor in (19) behaves at $k \to 0$ as $\frac{g(y)}{k}$ with some function $g(y)$. The contribution of the exceptional point $l = m = 0$, that is, $k = 0$ is described after transiting to the integration variables $k, y$, by the model integral

$$S_n = \frac{1}{2p} \text{Im} \int_0^{2p} g(y) \cdot \int_0^\infty e^{iktF(y)} h(k) dk \qquad (21)$$

where $F(y) = \frac{x}{t}\cos y + \frac{y}{t}\sin y - x_{0n}(y)$, and $h(k)$ is the patch function (neutralizator), that is, the finite infinitely differentiable function identically equal to a unit at smaller $k$. The asymptotics $S_n$ at $t \to \infty$ is not depending on the function $h(k)$ selection.

It the stationary points of the phase function (20) are limited starting from $l = m = 0$, then $F(y)$ has just simple zeroes. Moving the contour of integration over $y$ in (21) in the neighborhood of zeroes $F(y)$ to the upper semi-plane at $F'(y) > 0$ and to the lower semi-plane at $F'(y) < 0$, and after that integrating by parts in the integral over $k$ in (21), it is easy to prove that

$$S_n \approx \frac{1}{2pt} \int_0^{2p} \frac{g(y) dy}{F(y)} = \frac{1}{t} H\left(\frac{x}{t}, \frac{y}{t}\right) \qquad (22)$$

where the integral over $y$ is understood in the sense of principal value.

Now, we shall analyze the stationary points of the phase function $\Phi$. These points satisfy the equation

$$a = \frac{x}{t} = \frac{\partial w_n(l,m)}{\partial l}; \quad b = \frac{y}{t} = \frac{\partial w_n(l,m)}{\partial m}. \tag{23}$$

The set of points on the plane $(a,b)$, for which this system is resolvable, that is, $\Phi$ has some stationary points, can be naturally called the wave region. At $t \gg 1$ and $a = \frac{x}{t}$, $b = \frac{y}{t}$ located in this region the integral (5.2.19) is defined by these stationary points, and each stationary point $l_q, m_q$ has a coherent oscillating term in the asymptotics $G_n$

$$W_{n,q} \sim \frac{1}{t} A(l_q, m_q) \exp(i(l_q x + m_q y - w_n(l_q, m_q)t)). \tag{24}$$

The numerical calculation results and analytical evaluations indicate that in any case for $N(z)$, $V_1(z), V_2(z)$ getting closer to real expansions the wave region is bounded by two closed curves, i.e. by the leading and trailing edges. The leading edge is parametrically defined by the limit of formulas (23) at $k \to 0$ and $0 \le y \le 2p$; the trailing edge is defined at $k \to \infty$ and $0 \le y \le 2p$. When there are no flows the leading edge is the circumference of radius $C_n$ which is the maximum propagation velocity of the $n$-th mode, and the trailing edge is tightened to the origin of coordinates. At that the construction of the leading edge requires numerical calculations, while the trailing edge position is defined analytically.

In order to define the leading edge position we shall assume $w_n = k x_n(k,y)$, $F(z,y) = V_1(z) \cos y + V_2(z) \sin y$. Then $x_n$ is the eigenvalue of the spectral problem

$$(x-F)^2 \frac{\partial^2 j}{\partial z^2} + \left[ N^2(z) - k^2(x-F)^2 + \frac{\partial^2 F}{\partial z^2}(x-F) \right] j = 0 \tag{25}$$

$$j = 0 \quad (z = 0, -H)$$

moreover, $x_n(k,y)$ and the eigenfunctions $j = j_n(k,y,z)$ are expanded into series according to even exponents $k$:

$$x_n(k,y) = x_{0n}(y) + k^2 x_{1n}(y) + ...$$

(26)

$$j_n(k,y,z) = j_{0n}(z,y) + k^2 j_{1n}(z,y) + ...$$

Substituting these formulas into (24) and setting equal the coefficients with similar exponents $k$, we arrive at that $x_{0n}$ and $j_{0n}$ are the eigenvalue and eigenfunctions of the spectral problem

$$(x-F)^2 \frac{\partial^2 j}{\partial z^2} + \left(N^2(z) + \frac{\partial^2 F}{\partial z^2}(x-F)\right) j = 0; \quad j = 0 \ (z=0,-H).$$

From the equation resolvability condition for $j_{1n}$ we obtain the formula for $x_{1n}$, and so on.

If in (23) we assume $w_n = kx_n$, then at $k \to 0$ we get the following leading edge equations

$$\frac{x}{t} = a(y) = x_{0n}(y)\cos y - \frac{\partial x_{0n}}{\partial y}\sin y,$$

(27)

$$\frac{y}{t} = b(y) = \frac{\partial x_{0n}}{\partial y}\cos y + x_{0n}\sin y.$$

Let's clarify the geometrical meaning of this formula. The function $x_{0n}(y)$ is the maximal phase (and group) propagation velocity of the directed to $y$ plane wave $\exp ik[x\cos y + y\sin y - x_{0n}t] \cdot f(z)$. This wave edge, which at $t=0$ crossed the origin of coordinates $x = y = 0$, at some $t > 0$ shall travel the distance $tx_{0n}$ and takes the position $AA'$ (see Fig.1). The leading edge (27) of the mode $G_n$ is the envelope of the right lines $AA'$ with any directions of $y$: $0 < y < 2p$.

If $V_1(z)$ and $V_2(z)$ are little changed vertically $V_1 = V_{10} + d V_{11}(z)$; $V_2 = V_{20} + dV_{21}(z)$, where $d$ is small, then to an accuracy of values of the order $d^2$ the leading edge shall be a circumference expanding with velocity $C_n$ ,, (the maximal group velocity provided there are no flows) and a swept away (carried away by flows) with velocity equal to the mean value $V_1(z), V_2(z)$ taken with some weight.

In the neighborhood of the leading edge the function is expressed through the square of the Airy function

$$G_n \sim \frac{1}{t^{2/3}} A(y) \, Ai^2\left(\frac{h(y)r}{t^{1/3}}\right)$$

where $\rho$ is the distance from the observation point to the leading edge (the values $r > 0$ are corresponding to points located ahead of this edge); $y$ is the direction of the normal line lowered from the observation point to the edge, and the explicit expressions for the functions $A(y)$ and $h(y)$ shall not be provided here due to their awkwardness.

The trailing edge position, i.e. the limit $\frac{\partial w_n}{\partial l}$ and $\frac{\partial w_n}{\partial m}$ at $k \to \infty$ is calculated analytically. The trailing edge represents the same curve for all $n$. To construct it we shall take on the plane $a, b$ the velocity vector hodograph ($V_1(z), V_2(z)$), that is, the curve that is covered by the running point $a = V_1(z)$; $b = V_2(z)$ while $z$ changes from zero to $-H$. The trailing edge constitutes a convex hull of this curve. Fig.2 shows the oval $\Sigma$, which is the leading edge, the solid curve $PQ$ is the velocity hodograph, and the dashed line depicts the trailing edge.

Consider some point $P$ with the coordinates $x = r\cos j$, $y = r\sin j$, where $r$ is sufficiently large. On the plane $a, b$ at given $t$ this point is matched by some point $P_t$ with the coordinates $\frac{r}{t}\cos j$, $\frac{r}{t}\sin j$. At $t$ growing from zero to infinity, the point $P_t$ is moving to the origin of coordinates along the ray with inclination at $j$. When $P_t$ crosses the leading edge there occurs at $G_n$ a wave component, and this function begins to oscillate; when $P_t$ crosses the trailing edge the wave component disappears.

Of interest is the wave component asymptotics near the trailing edge. Relating model integrals are not reduced to certain known special functions, and yet make it possible to describe the qualitative peculiarities of the wave component $W_n$ behavior. Mentioned below are the corresponding results limited to the first mode $W_1$.

If the source is localized at certain horizon $z = z_0$, then, getting closer to the trailing edge, $W_1$ is exponentially tending to zero at any $z \neq z_0$. If, however, the source is vertically distributed over some interval $(z_-, z_+)$, that is, we examine a field representing the Green's function integral

$$\int_{z_-}^{z_+} G_1(t, x, y, z, z_0)\, h(z_0) dz_0,$$

then $W_1$ shall tend to zero, if $z$ is located outside the interval $(z_-, z_+)$. For the horizon $z$ located inside the interval $(z_-, z_+)$ the asymptotics $W_1$ depends on where the point $P_t = \left(\frac{x}{t}, \frac{y}{t}\right)$ with growing $t$ crosses the trailing edge.

This edge involves the hodograph $PQ$ arcs (these are $PM$ and $TQ$ arcs in Fig.2) and the convex hull segments ($LT$ and $QP$ segments in Fig.2). Let $P_t$ crosses the trailing edge at some point $M$ of the hodograph (at $P_t = P_t^1$ in Fig.2). This point is matched by some value $z = \tilde{z}$, for which $a = V_1(\tilde{z})$; $b = V_2(\tilde{z})$. While $P_t$ is getting closer to $M$ the field $W_1$ is concentrated at the horizon $z = \tilde{z}$, and the first mode group velocity is tending to $V_1(\tilde{z})$, $V_2(\tilde{z})$. In other words, at larger $t$ the field that is concentrated at the horizon $z = \tilde{z}$ has a zero velocity with respect to the medium (because its group velocity equals the flow velocity at this horizon).

We may say that the flows "break down" the field taking it to various horizons, and at each horizon the field is localized in that region which at $t = 0$ was at the origin of coordinates and after that with growing $t$, is whirled away with the flow velocity at this horizon. However, it doesn't occur at every horizon, but only at those that match the hodograph arcs contained in the convex hull (the arcs $PM$ and $TQ$ in Fig.2). If, however, $P_t = P_t^2$ crosses the trailing edge through the segments $LT$ and $QP$, then the field $W_1$ with getting closer to the trailing edge is exponentially decaying.

In summary we briefly shall consider the field of a source in uniform and straight motion with horizontal velocity $\mathbf{V}$, we shall assume that $|\mathbf{V}|$ is greater than the maximum group velocity. This field is not a difficult task to construct provided we know the Green's function. It is given as

$$W(\xi, z, z_0) = W_m + \sum_n W_n; \quad \xi_x = x - \mathbf{V}_x t; \quad \xi_y = y - \mathbf{V}_y t,$$

where $W_m$ is the contribution of the continuous spectrum, $W_n$ is the $n$-th mode, herewith $W_m$ is decaying at $|\xi| \to \infty$ faster than any $W_n$. Each of the modes far from the source $M$ features the leading edge ($MA$ and $MB$ in Fig.3) and the trailing edge ($MC$ and $MD$), before the rays $MA$, $MB$ and after rays $MC$, $MD$ a mode is exponentially small, in sectors $AMC$ and

*DMB* it oscillates, and in the neighborhood of rays *MA* and *MB* is given by the Airy function.

To find the location of edges we have to perform simple geometrical constructions. Let's construct the leading edge $\Sigma$ and trailing edge $S$ for the mode $G_n$ from the Green's function expansion and let down out of the point $O$ the source's velocity vector $OM = \mathbf{V} \cdot t$. Then, the source's location and the leading edge of the source's field shall be the tangential *MA* and *MB* to $\Sigma$, and the trailing edge shall be the tangential *MC* and *MD* to $S$. The far field of each mode in the neighborhood of it leading edge is expressed same as in the case of no flows by the Airy function [16-18].

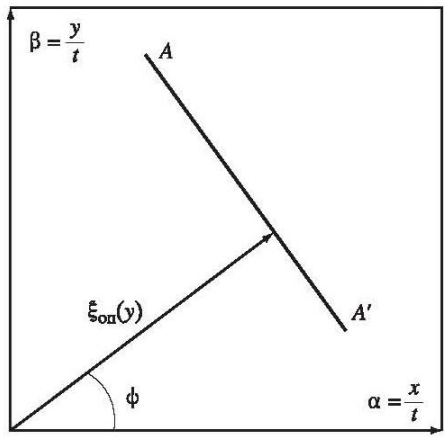

Fig.1 Wave front time-to-time evolution

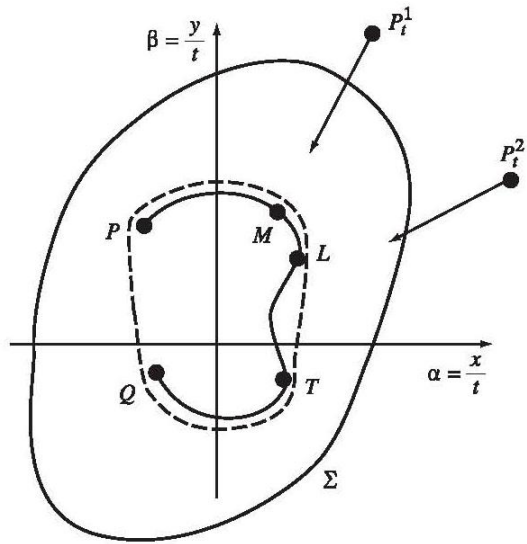

Fig.2 Wave fronts of internal gravity waves first mode generated by a stationary source

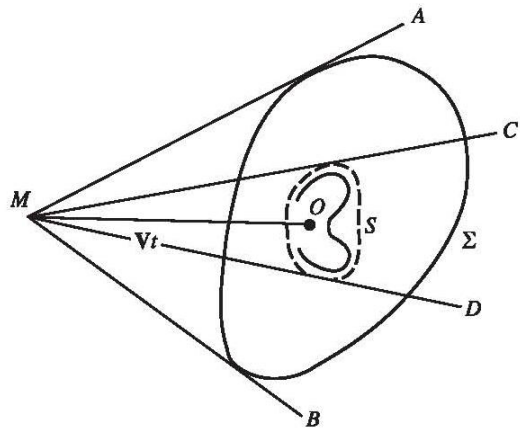

Fig.3 Wave fronts of internal gravity waves first mode generated by a moving source

.

**Conclusions.**

In paper we considered the fundamental problems of internal gravity waves dynamics in a stratified ocean with shear flows. The solution of this problem is expressed in terms of the Green's function and the asymptotic representations of the solutions are obtained. We construct the asymptotic representations of the far internal gravity waves generated by sources in stratified ocean with shear flows. Obtained in this paper results suggest that the parameters of shear flows and ocean density are independent on horizontal coordinates and time. Therefore, the aim of further research will be a study of internal gravity waves dynamics in the non-stationary, horizontally non-uniform stratified ocean with shear flows.

**Acknowledgments.**

The results presented in the paper have been obtained by research performed under projects supported by the Russian Foundation for Basic Research (No.11-01-00335a, No. 13-05-00151a), Program of the Russian Academy of Sciences "Fundamental Problems of Oceanology: Physics, Geology, Biology, Ecology" .